\documentclass[prl,aps,amssymb,superscriptaddress,showkeys,notitlepage,nobalancelastpage]{revtex4-1}
\usepackage[english]{babel}
\usepackage[utf8]{inputenc}
\usepackage{graphicx}
\usepackage{amsmath}
\usepackage{systeme}
\usepackage{bbold}
\usepackage{graphicx}
\usepackage{caption}
\usepackage[font=scriptsize]{caption}
\graphicspath{ {./images/} }
\usepackage[colorinlistoftodos, color=green!40, prependcaption]{todonotes}
\usepackage{amsthm}
\usepackage{mathtools}
\usepackage{physics}
\usepackage{xcolor}
\usepackage{graphicx}
\usepackage{adjustbox}
\usepackage{placeins}
\usepackage[T1]{fontenc}
\usepackage{lipsum}
\usepackage{csquotes}

\begin{document}
\title{Vectorial Bulk-Boundary Correspondence for 3D Photonic Chern Insulators}
  
\author{Chiara Devescovi}
    \email[]{chiara.devescovi@dipc.org}
     \affiliation{Donostia International Physics Center, 20018 Donostia-San Sebastián, Spain}
\author{Mikel García-Díez}
    \affiliation{University of the Basque Country (UPV-EHU), Bilbao, Spain}
    \affiliation{Donostia International Physics Center, 20018 Donostia-San Sebastián, Spain}
\author{Barry Bradlyn}
    \affiliation{Department of Physics and Institute for Condensed Matter Theory, University of Illinois at Urbana-Champaign, Urbana, IL, 61801-3080, USA}
\author{Juan Luis Mañes}
    \affiliation{University of the Basque Country (UPV-EHU), Bilbao, Spain}
\author{Maia G. Vergniory}
    \email[]{maiagvergniory@dipc.org}
     \affiliation{Donostia International Physics Center, 20018 Donostia-San Sebastián, Spain}
    \affiliation{Max Planck Institute for Chemical Physics of Solids, Dresden D-01187, Germany}
\author{Aitzol Garc\'{i}a-Etxarri}
    \email[]{aitzolgarcia@dipc.org}
    \affiliation{Donostia International Physics Center, 20018 Donostia-San Sebastián, Spain}
    \affiliation{IKERBASQUE, Basque Foundation for Science, Maria Diaz de Haro 3, 48013 Bilbao, Spain}

\date{\today} 

\begin{abstract}

In 2D Chern insulators (2D CI), the topology of the bulk states is captured by a topological invariant, the Chern number. The scalar bulk-boundary correspondence (sBBC) relates the change in Chern number across an interface with the number of 1D chiral edge modes at the interface. However, 3D Chern insulators (3D CI) can be characterized by a  Chern vector $\mathbf{C}=(C_x,C_y,C_z)$ and a more general vector bulk-boundary correspondence (vBBC) is needed to correctly predict the propagation of the surface modes. 
In this work the possible interfaces between 3D photonic CIs are explored, focusing on possible changes in Chern vector orientation.

To formulate a 3D vBBC, a link is derived between the Chern vector discontinuity across an interface and the winding of the surface equifrequency loops on the boundary. Lastly, it is demonstrated how to correctly predict the number and the propagation direction of topological photonic surface modes in 3D CIs.
\end{abstract}

\maketitle
\begin{center}
\large\textbf{Introduction}
\end{center}
\normalsize

A 3D Chern insulator (3D CI) is a Time-Reversal Symmetry (TRS) broken topological phase characterized by a Chern vector $\mathbf{C}=(C_x,C_y,C_z)$  \cite{xu2020high,vanderbilt2018berry,halperin1987possible,hughes2011inversion,haldane2004berry,devescovi2021cubic} that can support anomalous surface states (SS) on surfaces parallel to the Chern vector. Surface states are usually considered to be unidirectional in the following sense: The (conserved) component of the group velocity normal to the magnetization axis (i.~e. the Chern vector direction) has a well-defined sign and surface states cannot back-scatter along this specific direction.

In 2D, the Chern vector is always fixed along the axis of the reduced dimensionality, i.e. orthogonal to the plane of the system. Therefore, it can be regarded as a scalar quantity: the Chern number $C$, which characterizes the bulk topology of 2D CIs \cite{haldane1988model,hasan2010colloquium,blanco2020tutorial}. In this case, a 'scalar' version of the bulk-boundary correspondence (sBBC) can be defined, to connect the bulk topology to the number of boundary modes \cite{halperin1982quantized,qi2011topological}. According to sBBC in 2D CIs, an interface between two systems with Chern numbers $C_1,C_2$ has $n_e=|{{C}_1-{C}_2}|$ protected chiral edge states. This means that chiral edge states can only appear in presence of a discontinuity of Chern numbers across the interface, i.e. $C_1\neq C_2$ \cite{wang2008reflection,wang2009observation,raghu2008analogs,haldane2008possible}. 

On the contrary, the interface between two 3D CIs can have Chern vectors $\mathbf{C}_1$ and $\mathbf{C}_2$ that need not be parallel/anti-parallel to each other \cite{vanderbilt2018berry,kohmoto1993quantized}. This leads to vectorial aspects of the BBC which must be taken into account in order to correctly define a 'vectorial' bulk-boundary correspondence (vBBC). 

In the broader context of topological insulators, the literature has discussed the vectorial aspects of BBC mostly in relation to 3D quantum spin-Hall fermionic insulators \cite{fu2007topological,moore2007topological,roy2006three}. Such systems are characterized by vectors of $Z_2$ invariants and rely on time-reversal symmetry (TRS). However, the photonic 3D CIs studied in this manuscript break TRS and display non-zero Chern vectors, requiring a suited vectorial BBC. While scalar bulk-boundary correspondence is a well-established concept in 2D photonic crystals with broken TRS \cite{hafezi2014measuring,wang2009observation,wang2008reflection,raghu2008analogs,haldane2008possible}, less is known in photonic 3D CIs, recently demonstrated in \cite{devescovi2021cubic},  which is what motivated our work. We believe that this limited knowledge is related to the fact that, in condensed matter systems, the 3D Quantum Hall effect is usually discussed in layered systems \cite{bernevig2007theory,yang2013topological,cho1997transport,meir1998quantum,metzler1998spectral, zheng2020thouless,druist19982d,balents1996chiral}. Since layered systems have a preferred axis, and since the magnetic field needs to be along this axis, the vectorial nature doesn't come up.  On the contrary, the possibility of orienting Chern vectors in space demonstrated in photonic crystals \cite{devescovi2021cubic}, open up the possibility of constructing domain walls between different orientations, and thus a definition of a vectorial BBC is required.

Consider, for example, a planar interface with Chern vectors orthogonal to each other, i.e. $\mathbf{C}_1\cdot \mathbf{C}_2=0$. Different questions can arise: Do individual components of the Chern vector contribute in a linear independent way to surface modes? How do multiple photonic surface modes hybridizee with each other with respect to the different orientations of the Chern vector? Is there an easy way of counting surface modes or predicting their direction? Finding answers to these questions is a challenging theory exercise, lacking of a 2D analogy.  This is even more relevant if we consider that the surface of a 3D CI exhibits equifrequency ``Fermi'' loops \cite{vanderbilt2018berry,devescovi2021cubic}, i.e. surface modes with a dispersion formed by closed equi-frequency lines that wind around nontrivial cycles of the surface Brillouin zone (BZ). Understanding the connection between the Fermi loop connectivity and Chern vector change at the interface is one of the goals of this work.

Beyond theoretical concerns, constructing interfaces with Chern vectors of different orientations is relevant for practical 
applications. 3D photonic CI interfaces are a potential platform for unidirectional optical channels protected from backscattering\cite{devescovi2021cubic}. Furthermore, photonic crystal architectures with multiple Chern vectors of different orientation could enable new ways to control light propagation. As an example, a cubic arrangement of magnetized 3D CI panels around an inert core, with Chern vector pointing inwards (e.g. fixing a 3D $+C_x$I on a left $\hat{\mathbf{x}}$ panel) was originally proposed as a magneto-electric (ME) coupler in electronic systems \cite{vanderbilt2018berry}, as an alternative route to Chern wrappers \cite{mogi2017tailoring,mogi2017magnetic,wang2015quantized}: a photonic equivalent still needs to be discussed. 

In this work, in the context of photonic crystals, we investigate vectorial aspects of bulk-boundary correspondence at the interface of 3D CIs with different Chern vectors $\mathbf{C}_1 \neq \mathbf{C}_2$. In particular, we analyze the emerging surface states and show what occurs when the Chern vectors across the boundary are not parallel to each other. The analysis of interfaces where Chern vectors have a different orientation in space allows us to promote the sBBC from 2D to a 'vectorial' bulk-boundary correspondence (vBBC) for 3D CIs.

\begin{center}
\large\textbf{Methods}
\end{center}
\normalsize

\subsection{Bulk: 3D CIs with orientable Chern vectors} \label{sec:setup}

Photonic 3D CIs with orientable Chern vectors constitute the building blocks for the photonic interfaces and are constructed according to the four-steps supercell modulation strategy proposed in \cite{devescovi2021cubic}: (1) design of the photonic crystal unit, (2) Weyl dipole formation, (3) supercell band-folding and (4) 3D Chern gap opening by Weyl points annihilation.

Each of these steps is graphically visualized in Figs. 1-4 and summarized in the following subsections 2.1.1-2.1.5. Each figure shows the unit cell, the irreducible Brillouin zone and the corresponding band-structure, with frequencies $f$ in reduced units, $|a|$ being the scale invariant lattice parameter and $c$ the speed of light.

\vspace{0.5cm}
\subsubsection{Design of the photonic crystal unit}
As a first step, we consider a photonic crystal (space group No.~224) with a unit cell containing four dielectric rods of radius $r_0=0.1$ directed along the main diagonals of a cube and meeting at its center. The dielectric material is described by a diagonal permittivity tensor, $\varepsilon_{TRS}=\varepsilon\mathbb{1}_3$, with $\varepsilon=16$, and by unit magnetic permeability $\mu=\mathbb{1}_3$.  The photonic band-structure presents a three-fold degeneracy at the high symmetry point $\mathbf{R}=\frac{2\pi}{|a|}(1/2,1/2,1/2)$, as shown in Fig.1.

\begin{figure}[h!]
\centering
\includegraphics[width=86mm]{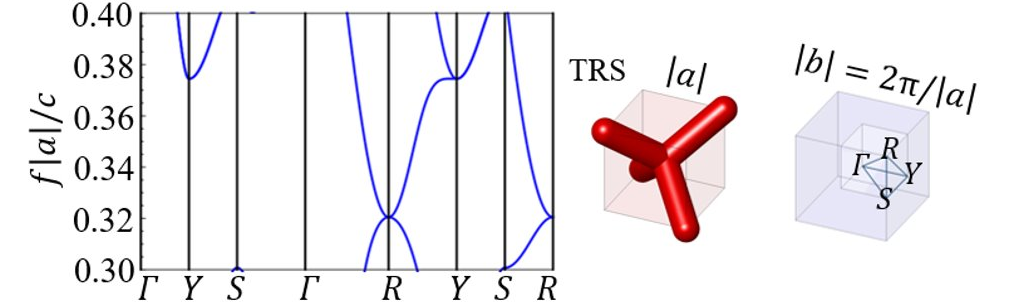}
\captionsetup[figure]{font=small}
\caption{Photonic crystal of rods with radius $r_0=0.15$ and dielectric constant $\varepsilon=16$ at TRS and unit lattice parameter $|a|=1$). Three-fold degeneracy at $\mathbf{R}$ between the lowest three bands.}
\label{s1}
\end{figure}

\subsubsection{Weyl dipole formation}
In the second step, we magnetize the crystal to split this trifold degeneracy into Weyl points: the 3D CI phase will be constructed by annihilating these Weyl points. This can equivalently be achieved by exploiting the gyro-electric response of a photonic crystal \cite{yang2017weyl} or via gyro-magnetic materials \cite{wang2008reflection}, by introducing off-diagonal imaginary elements in the permittivity or the permeability tensor, respectively. For example, for a desired magnetization along $\hat{\mathbf{z}}$,  a gyro-electric response can be simulated as
\begin{equation}\label{eq:gyro}
{\varepsilon}_{\eta_z}=\left(\begin{array}{ccc}
\varepsilon_\perp & i\eta_{z} & 0 \\
-i \eta_{z} & \varepsilon_\perp & 0 \\
0 & 0 & \varepsilon
\end{array}\right),
\end{equation}
where $\eta_{z}=\eta_{z}(B_z)$ the bias-dependent gyro-electric parameter and $\varepsilon_{\perp}=\sqrt{\varepsilon^2+\eta_z^2}$. Similarly, for a desired magnetization along $\hat{\mathbf{z}}$,  a gyro-magnetic response can be simulated as:
\begin{equation}\label{eq:gyro}
{\mu}_{\rho_z}=\left(\begin{array}{ccc}
\mu_\perp & i\rho_{z} & 0 \\
-i \rho_{z} & \mu_\perp & 0 \\
0 & 0 & \mu
\end{array}\right)
\end{equation}
where $\rho_{z}$ is the gyro-magnetic parameter, $\mu$ is the permeability in the absence of remnant magnetization, and $\mu_{\perp}=\sqrt{\mu^2+\rho_{z}^2}$; other directions can be obtained by orthogonal rotations. 
For numerical convenience, we chose the first approach. However, we stress that both approaches are equivalent from a topological point of view \cite{haldane2008possible,raghu2008analogs,devescovi2021cubic,wang2008reflection}.
As a consequence of the magnetization, the trifold degeneracy splits into a Weyl dipole oriented along the same direction. The Weyl splitting can be controlled by tuning the parameters of the crystal and the intensity of the gyrotropic response. Specifically, we want the Weyl points to be located at fractional distances of the Brillouin zone, i.e. at positions $\mathbf{K}_{1,2}=\mathbf{R}\pm \frac{\mathbf{X}_i}{N_W}$ where $N_W\in \mathbb{N}$ and $N_W>1$ where $\mathbf{X}_i=\frac{\mathbf{b}_{i}}{2}$ and $\mathbf{b}$ are the reciprocal lattice vectors. For $\eta_z=16$ the Weyl points separation is half the BZ, as shown in Fig.2. The formation of a Weyl dipole is important because, from its annihilation, will result the 3D CI phase (as shown in Sec 2.1.4). Moreover, the orientation of the Weyl dipole determines the orientation of the resulting Chern vector (as shown in Sec 2.1.5).

\begin{figure}[h!]
\centering
\includegraphics[width=86mm]{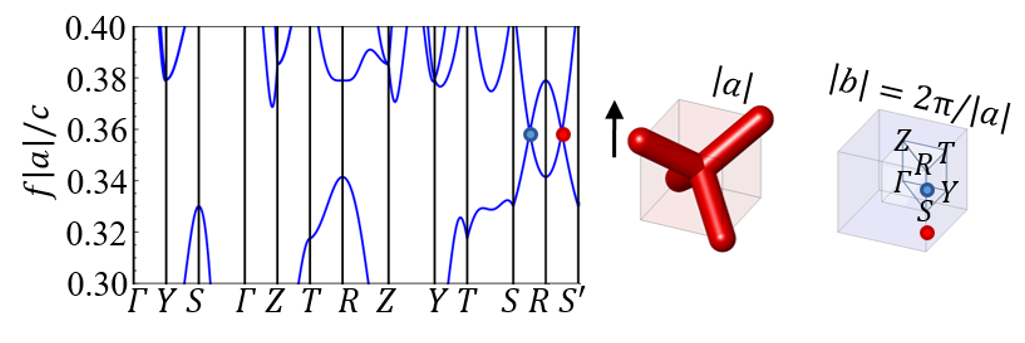}
\captionsetup[figure]{font=small}
\caption{Magnetization of the crystal implemented via a gyrotropic response with $\eta_z=16$, which results in the formation of a Weyl dipole. The parameters are adjusted in order to split the Weyl points of approximately half the BZ. }
\label{s2}
\end{figure}

\subsubsection{Supercell band-folding}
As a third step, we fold the BZ by creating multi-fold  ($N>1$) supercells; this is achieved by replicating the original unit cell either in a cubic supercell of dimensions $(N,N,N)$. As a consequence, the Weyl points overimpose on top of each other in the new reduced BZ, forming a four-old degeneracy, as in Fig.3. In this manuscript, to keep simulations affordable, we fix $N=N_W=2$. 

\begin{figure}[h!]
\centering
\includegraphics[width=86mm]{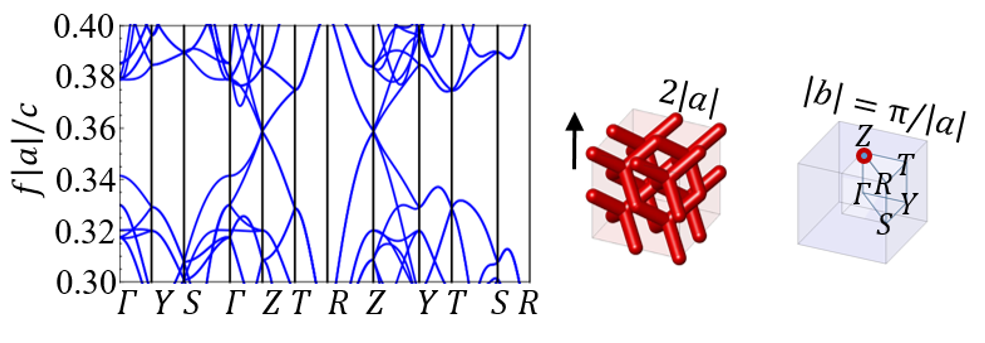}
\captionsetup[figure]{font=small}
\caption{Supercell folding of the bands on a $N=2$ cubic supercell. }
\label{s3}
\end{figure}

\subsubsection{ 3D Chern gap opening by Weyl points annihilation}
As a fourth and last step, we couple the opposite-charge Weyl points and gap the artificial fourfold degeneracy opposite-charge Weyl points by spatially modulating the crystal geometry with a periodicity commensurate with the supercell. In this specific setup, we locally vary the radius of the cylinders through the entire supercell, from their original $r_0$ radius to the new local one $r(x,y,z)$ according to a relation of the type: $
          \Delta r(x,y,z)=r(x,y,z)-r_0
          =r_m[\cos(2 \pi x/N|a|) + \cos(2 \pi y/N|a|) + \cos(2\pi z/N|a|)]$
where $r_m$ controls the intensity of the modulation. Here $r_m/r_0=1/20$.
As a result (see Fig.4), a 3D Chern gap is opened and the system displays a non-zero Chern vector.

\begin{figure}[h!]
\centering
\includegraphics[width=86mm]{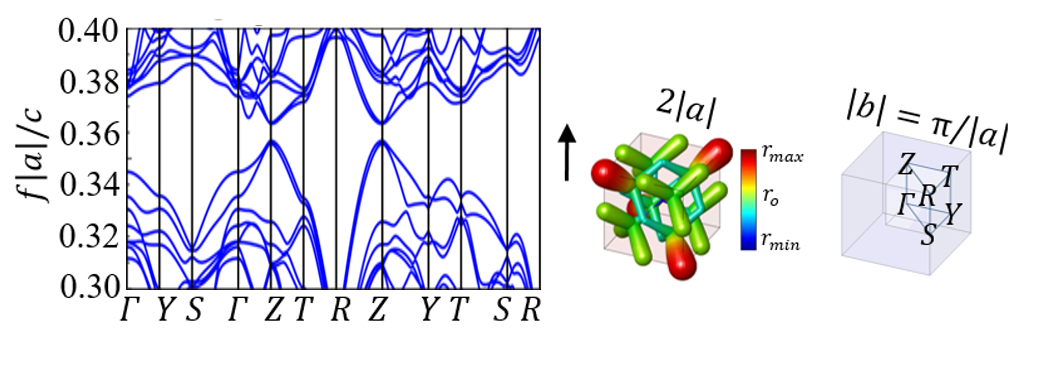}
\captionsetup[figure]{font=small}
\caption{ 3D Chern gap opening. Annihilation of Weyl points through a supercell modulation with $N=2$ and $r_m=r_0/20$. A colorbar is associated to the local radius of the cylinders, where $r_{max}=r_0+r_m$ and $r_{min}=r_0-r_m$. As a result, a 3D Chern gap is opened and the resulting system displays a unit Chern vector oriented along $z$.}
\label{s4}
\end{figure}

\subsubsection{Orientable Chern vectors}
The Chern vector for our 3D CI cubic model is fixed by the magnetization direction of Eq.1, as shown in Fig.5. Therefore, by choosing gyrotropic materials, it is possible to arrange magnetized samples in the desired configuration to construct domain-wall interfaces. 

\begin{figure*}[ht]
\centering
\includegraphics[width=158mm]{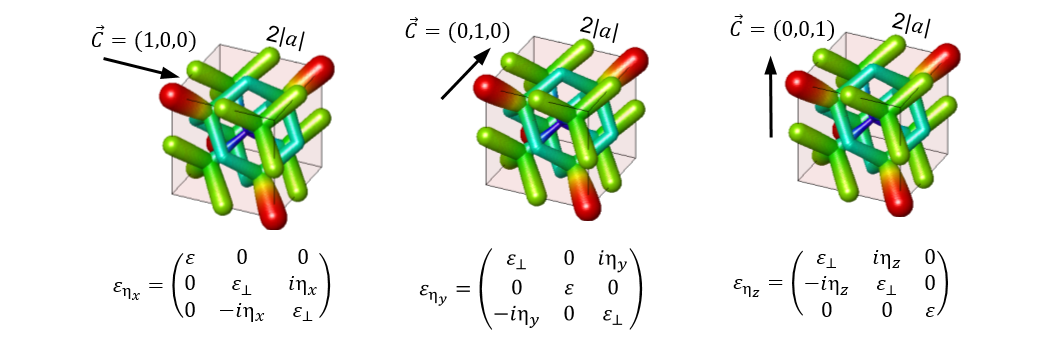}
\captionsetup[figure]{font=small}
\caption{Orientability of the Chern vector along the gyrotropic axis.}
\label{s4}
\end{figure*}

\subsection{Surface: domain-wall planar interfaces} \label{sec:supercells}
To analyze planar interfaces at the boundary of different 3D CIs, we construct 1D supercells and obtain topological slabs with normal vector oriented along $\hat{\mathbf{x}}$. In particular, we analyze the following 3D CI/3D CI interfacing configurations:
\begin{enumerate}
    \item $\mathbf{C}_1=C_{z_1}\hat{\mathbf{z}}$ and $\mathbf{C}_2=C_{z_2}\hat{\mathbf{z}}$, with $C_{z_1}=-C_{z_2}=C_{z}$: anti-parallel Chern vectors, both parallel to the interface plane. We call this configuration anti-ferromagnetic $C_z$I/$C_z$I.
    \item $\mathbf{C}_1=C_x\hat{\mathbf{x}}$ and  $\mathbf{C}_2=C_z\hat{\mathbf{z}}$: Chern vectors orthogonal to each other, one parallel to the interface plane, the other orthogonal to the interface plane. We call this configuration $C_x$I/$C_z$I.
    \item $\mathbf{C}_1=C_y\hat{\mathbf{y}}$ and  $\mathbf{C}_2=C_z\hat{\mathbf{z}}$: Chern vectors orthogonal to each other, both parallel to the interface plane.  We call this configuration $C_y$I/$C_z$I.
\end{enumerate}

There are two additional interface conditions: (4) The case for which $\mathbf{C}_1=0$ and $\mathbf{C}_2=C_{z_2}\hat{\mathbf{z}}$, i.e. a non-null Chern vector parallel to the interface with a trivial insulator; (5) The case where $\mathbf{C}_1=C_{z_1}\hat{\mathbf{z}}$ and $\mathbf{C}_2=C_{z_2}\hat{\mathbf{z}}$, with $C_{z_1}=C_{z_2}=C_{z}$, i.e. identical Chern vectors, both parallel to the interface plane. 
Case (4) was already extensively studied in Ref. \cite{devescovi2021cubic} and it here is referred to as Trivial/$C_z$I configuration. Case (5) will not have protected chiral modes by the scalar bulk-boundary correspondence, and so will not be of further interest here. However such interfaces may be interesting from the point of view of ``higher order'' topology \cite{wieder2020axionic}, and will be explored in a future work \cite{devescovi2022hinge}.

For each interface (1)--(3), we characterize the number and propagation direction of the emerging surface states and investigate the vectorial aspects of BBC using a combined real-reciprocal space analysis. First, we analyze the surface states in the frequency domain, showing the SS energy sheets as a function of surface momenta $k_y,k_z$  on the (100) plane. We then look at midgap equifrequency cuts to extract Fermi loops. Second, we analyze the direction of the associated group velocity (or Poynting vector) and the spatial localization of the fields on the boundary. Note that, for numerical convenience, we analyze only systems with unit Chern vectors, i.e. where $C_{x},C_{y},C_{z}$ are either $1$ or $-1$. Extending our considerations to larger Chern vectors would require larger supercells that are computationally too challenging for our resources. Nevertheless, unit chern vectors allow us to derive general principles of the vBBC that generalize to Chern vectors of arbitrary magnitude and direction.

\begin{center}
\large\textbf{Results}
\end{center}
\normalsize

\textbf{Figure 6}, \textbf{7} and \textbf{8} summarize the surface states analysis for the $C_z$I/$C_z$I, $C_x$I/$C_z$I and $C_y$I/$C_z$ configurations, respectively. Each figure displays: in panel a, the orientation of the Chern vectors as shown by arrows together with the unit cell of the 3D Chern photonic crystals; in panels b-c, the band dispersion, for the surface states (light blue) and for the nearby bulk bands (orange); in panel d, the equi-frequency Fermi loops in momentum space, with arrows indicating the direction of the group velocity (or, equivalently, the Poynting vector \cite{mikki2009electromagnetic}). Red and green lines label the localization of surface modes, respectively, on the left or right side of the interface. 

\begin{center}
\large\textbf{Discussion}
\end{center}
\normalsize

\subsection{Case $C_z$I/$C_z$I} \label{sec:z_z}
In the case where the Chern vectors are anti-parallel, along $\hat{\mathbf{z}}$, and both parallel to the interface surface, the configuration is similar to an anti-ferromagnetic arrangement of opposite Chern numbers for a 2D system in the (001) plane \cite{kim2020recent}.
Therefore, for this simple case, we can expect sBBC to naturally extend to 3D. Specifically, when $C_{z_1}=-C_{z_2}=C_z$, we have $|{2{C_z}}|$ chiral boundary modes. 
Indeed, the configuration with anti-parallel unit Chern vectors $C_{z_1}=-C_{z_2}=1$ results in two co-propagating SS on each side of the supercell slab, as in Figure 6. Note that the conserved propagation component for these surface states coincides with the propagation direction of the chiral edge modes that would live in the 2D analog system. 
The 2D analogy employed here works because 3D CIs can be interpreted, in a layer construction picture \cite{xu2020high,elcoro2020magnetic}, as a stack of 2D CIs.

\begin{figure*}[h!]
\centering
\includegraphics[width=178mm]{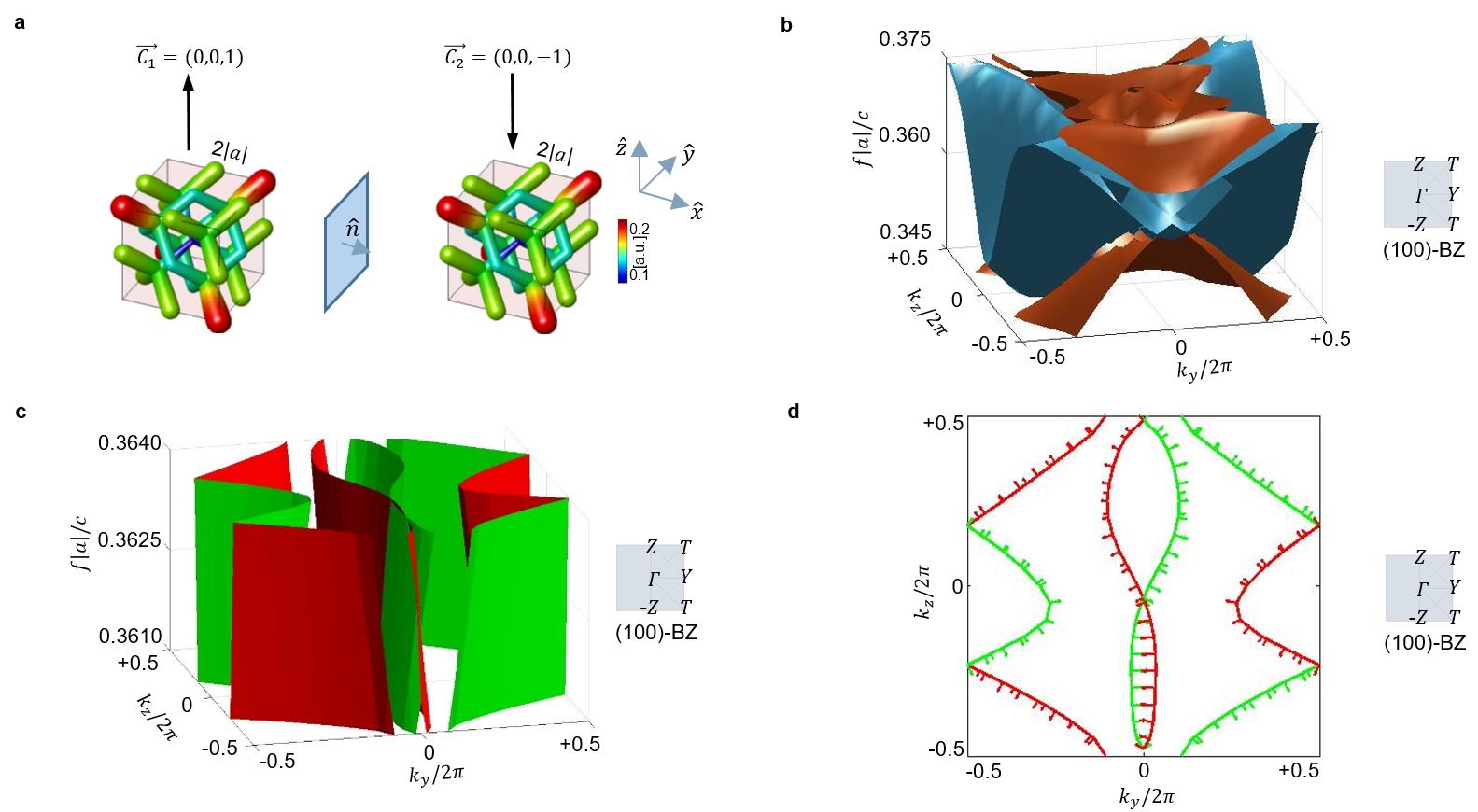}
\captionsetup[figure]{font=small}
\caption{$C_z$I/$C_z$I interface. $C_{z_1}=-C_{z_2}=1$. Two co-propagating surface states on each side of the slab.  a) shows the orientation of the Chern vectors (by arrows) together with the photonic crystal unit cell on each side of the interface. The color code indicates the value of the local radius of the cylinders which constitute the rods of the photonic crystal. The radii of the dielectric rods are locally modulated in order to induce a supercell modulation effect able to open up the 3D Chern gap in the underlying crystals, as explained in the SI Sec.1. b) shows the surface (light blue) and bulk (orange) state dispersion as a function of $k_y$ and $k_z$ for frequencies near the topological gap. c) shows the surface state dispersion over a narrow frequency range in the bulk band-gap.  Green (red) colors denote states on the right (left) interface. d) shows an equi-frequency cut of the surface states. The direction of the Poynting vector is denoted by arrows.}
\label{zz}
\end{figure*}

\subsection{Case $C_x$I/$C_z$I} \label{sec:x_z}
In 3D CIs, unidirectional surface states can emerge only on the planes parallel to the magnetization \cite{xu2020high,elcoro2020magnetic,vanderbilt2018berry}. Therefore, we can expect the 3D CIs with Chern vector normal to the interface plane (the $C_x$I side) not to contribute to the spectral flow on the boundary. As a confirmation of this, we observe that both the SS dispersion and the Fermi loop displayed in Figure 7 constitute an adiabatic deformation of those found on the 3D $C_z$I/trivial interface of Ref. \cite{devescovi2021cubic}. In particular, the conserved propagation component is determined by $C_z$ only, yielding a surface mode with velocity along $k_y$. Effectively, the $C_x$I side does not alter the number of surface modes on the $\hat{\mathbf{x}}$ interface. Therefore, the propagation and the spectral properties of the surface modes do not change even when reversing the sign of $C_x$. Note, however, that the presence of a band-gap in the $C_x$I side (in this case a Chern gap) is fundamental for it to behave as an insulator. In conclusion, due to the orientation of its Chern vector,  the $C_x$I side acts as a 'trivial' insulator from the perspective of surface states, even though the Chern vector itself is not zero. A $C_x$I/$C_y$I configuration for the same $\hat{\mathbf{x}}$ boundary can be treated very similarly. In such a case, the Fermi loops would be rotated in the $k_y-k_z$ plane relative to Figure 7, propagating along $z$. These facts will be exploited in the following subsection 4.3.

\begin{figure*}[h!]
\centering
\includegraphics[width=178mm]{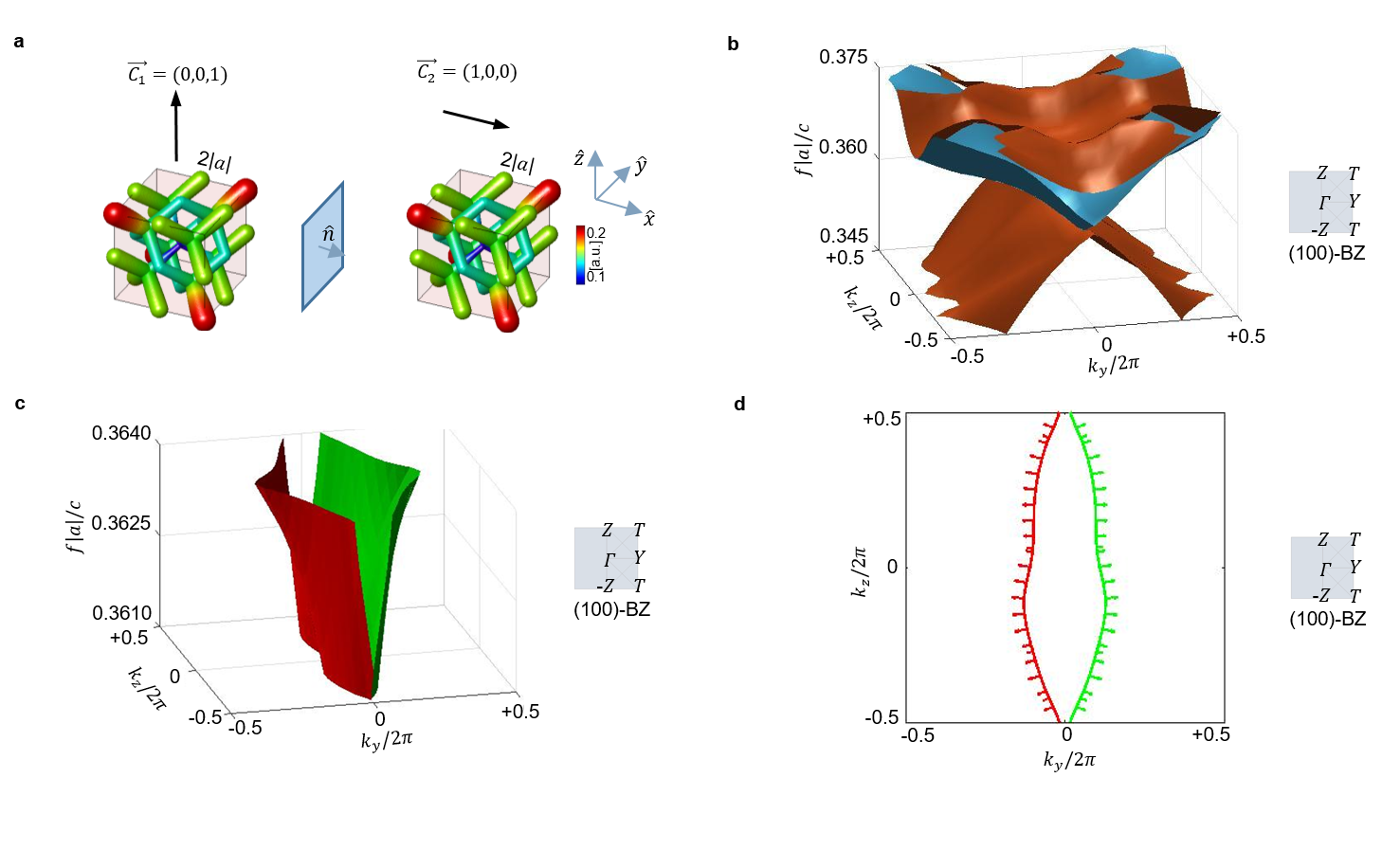}
\captionsetup[figure]{font=small}
\caption{3D $C_z$I/3D $C_x$I interface. $C_z=1$ and $C_x=-1$. Identical when  $C_z=1$ and $C_x=1$. A single unidirectional surface state: lack of contribution from the Chern vector normal to the interface plane. a) shows the orientation of the Chern vectors (by arrows) together with the photonic crystal unit cell on each side of the interface. The color code indicates the value of the local radius of the cylinders which constitute the rods of the photonic crystal. The radii of the dielectric rods are locally modulated in order to induce a supercell modulation effect able to open up the 3D Chern gap in the underlying crystals, as explained in the SI Sec.1.
b) shows the surface (light blue) and bulk (orange) state dispersion as a function of $k_y$ and $k_z$ for frequencies near the topological gap. c) shows the surface state dispersion over a narrow frequency range.  Green (red) colors denote states on the right (left) interface. d) shows an equi-frequency cut of the surface states.  The direction of the Poynting vector is denoted by an arrow.}
\label{xz}
\end{figure*}

\subsection{Case $C_y$I/$C_z$I} \label{sec:y_z}
In this configuration, the Chern vectors are orthogonal to each other, and both parallel to the interface plane. This is a fully 3D configuration of Chern vectors, and so a 2D scalar analogy cannot be applied. To understand the correct counting of surface modes and their propagation on the boundary, we need to consider how SS can hybridize with each other. Following what was observed in the previous subsections, both the $C_y$I and $C_z$I side should contribute to surface modes with, respectively, propagation along $k_z$ and $k_y$.

However, in the plot of Figure 8 we observe a single surface state propagating along $(0 1 1)$, i.e. along the vector sum of the unit Chern vectors. This can be understood as follows. In a layer construction picture, the surface modes can never scatter along the propagation direction \cite{vanderbilt2018berry}, but they can hybridizee along the Chern vector direction. This reasoning will be supported by topological considerations and by a simple model in section 4.4.
Therefore, if we superimpose the plot of Figure 7d, i.e. for the $C_z$I side, with a 90 degrees tilted version of it, i.e. for the $C_y$I side, and if we allow only states on the same side of the sample to hybridizee, we obtain the Fermi loop connectivity of Figure  8d. Therefore, Figure  9 can now be interpreted via a linear combination of unit Chern vectors, as summarized in Figure 9b and 9c.

\begin{figure*}[h!]
\centering
\includegraphics[width=178mm]{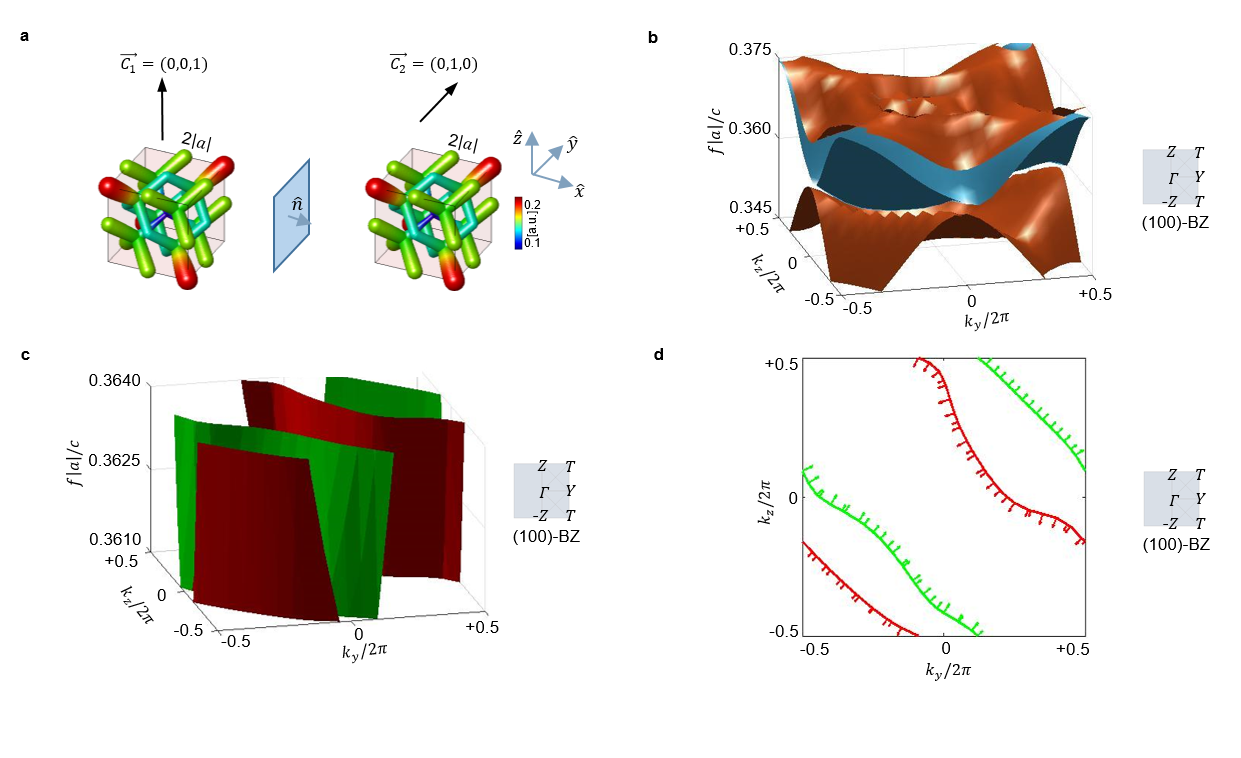}
\captionsetup[figure]{font=small}
\caption{3D $C_z$I/3D $C_y$I interface. A single surface state propagating along $(0 1 1)$, i.e. along the vectorial sum of the unit Chern vectors.
a) shows the orientation of the Chern vectors (by arrows) together with the photonic crystal unit cell on each side of the interface. The color code indicates the value of the local radius of the cylinders which constitute the rods of the photonic crystal. The radii of the dielectric rods are locally modulated in order to induce a supercell modulation effect able to open up the 3D Chern gap in the underlying crystals, as explained in the SI Sec.1. b) shows the surface (light blue) and bulk (orange) state dispersion as a function of $k_y$ and $k_z$ for frequencies near the topological gap. c) shows the surface state dispersion over a narrow frequency range. Green (red) colors denote states on the right (left) interface. d) shows an equi-frequency cut of the surface states.  The direction of the Poynting vector is denoted by an arrow.}
\label{yz}
\end{figure*}

\subsection{Topology of Fermi loops and vBBC} \label{sec:loops}
From our previous considerations,  it is possible to gain more insight into the topological properties of the Fermi loops and infer some general statements about vBBC for the case of a 3D CI. 

In general, it is well known that Fermi loops can split in disjoint chiral partners located on opposite sides of the interface slab, \cite{devescovi2021cubic,vanderbilt2018berry}, which is reproduced here as well, as confirmed by the fields localization analysis. When residing on the same side of the sample, overlapping Fermi loops can hybridize giving rise to a single, continuously connected Fermi loop, as sketched in Figures 9b--9e. The hybridization results in a local change of Poynting vector, for each specific momentum component. However, the winding of the resulting loop around the surface BZ is the same as that of the individual loops before hybridization. 

To see how this works quantitatively, we can consider a simple model for the interface between a $C_z=1$ and a $C_y=1$ 3D Chern system, again with the interface normal to the $\mathbf{\hat{x}}$ direction. 

Without any hybridization, we will have two chiral surface modes propagating on the interface: $\Delta C_z=1$ across the interface implies the existence of a state whose Hamiltonian can be written as $H_z=v_z k_y$, describing a chiral mode propagating in the $\mathbf{\hat{y}}$ direction. Similarly, $\Delta C_y=-1$ at the interface implies the existence of a mode with low energy effective Hamiltonian $H_y=v_yk_z$, describing a chiral mode propagating in the $\mathbf{\hat{z}}$ direction. The model employed here represents an effective model for the dispersion of an unidirectional mode on the surface, with broken TRS and at lowest order in $\mathbf{k}$.  Ignoring any hybridization between these modes, we can write the combined surface effective Hamiltonian as:

\begin{equation}
    H_0 =\begin{pmatrix}
    v_zk_y & 0 \\
    0 & v_yk_z 
    \end{pmatrix}.
\label{eq:surfaceham0}
\end{equation}
The effective Hamiltonian description allows us to find an expansion in powers of the wave vector $\mathbf{k}$ of the photonic energy bands $\omega$,  able to replicate the photonic modes dispersion obtained by numerically solving the Maxwell equations. 
The constant frequency contours of $H_0$ consist of one horizontal and one vertical line at each point in the surface Brillouin zone. The two modes at frequency $\omega$ intersect at $(k_y,k_z) = (\omega/v_z, \omega/v_y)$ Furthermore, there is spectral flow from negative to positive frequency in $H_0$, in that we can continuously follow each surface band from $-\infty$ to $+\infty$ in frequency. However, the twofold degeneracy of states at $(k_y,k_z) = (\omega/v_z, \omega/v_y)$ is generically unstable to surface perturbations. To lowest order, we can model such a perturbation as a constant off-diagonal term added to the effective surface Hamiltonian \eqref{eq:surfaceham0},
\begin{equation}
    H=H_0+V = \begin{pmatrix}
        v_zk_y & a \\
    a^* & v_yk_z
    \end{pmatrix},
\end{equation}
where $|a|$ parametrizes the strenght of the hybridization between modes. The eigenfrequencies of $H$ are given by
\begin{equation}
    \omega_{\pm} = \frac{1}{2}(v_zk_y+v_y k_z) \pm \frac{1}{2}\sqrt{(v_zk_y-v_xk_z)^2+4|a|^2}.
\end{equation}
We see immediately that $\omega_+\neq\omega_-$ for any $(k_y,k_z)$, so there is no longer degeneracy between states. The constant energy contours are now hyperbolic, just as in Figure 9. Although there is a gap between $\omega_+$ and $\omega_-$ at each momentum in the surface Brillouin zone, however, there is still spectral flow from the valence states to the conduction states in frequency and the edge state dispersion extends throughout the entire gap. To see this, we can consider the line $t=v_zk_y = v_yk_z$ in the surface Brillouin zone. Along this line, we have $\omega_\pm = t\pm|a|$; Thus both $\omega_+$ and $\omega_-$ interpolate from $-\infty$ to $+\infty$ as a function of $t$, consistent with the bulk-boundary correspondence.

To generalize this hybridization picture to interfaces with larger Chern vectors, we can employ a circuit- or network-type analogy, where the loop states are viewed as wires. If we consider a $\hat{\mathbf{x}}$ interface with $M$ loops directed along $k_y$ and $N$ loops directed along $k_z$, we can picture the BZ as sort of a black box with $M$ leads on the left and right edges, and $N$ leads on the top and bottom edges. If we want to wire up this box so that current can flow through all the leads, the only way to do this with no intersections is to have a single wire that winds from left to right $M$ times and from botton to top $N$ times. Since the surface BZ has the topology of a torus, we have $(T_1, T_2)=(M,N)$, where $T_1,T_2$ denote the winding around the two handles of the surface BZ torus. Therefore, given a difference of Chern numbers equal to:
\begin{equation}
    \Delta\mathbf{C}=\mathbf{C}_1-\mathbf{C}_2=(C_{x_1}-C_{x_2},-M,N),
\end{equation} 
we can expect $M=C_{y_1}-C_{y_2}$ loops directed along $k_y$ and $N=C_{z_1}-C_{z_2}$ loops directed along $k_z$, since,  as observed in subsection 4.2, the individual values of $C_x$ do not contribute to any mode on a $\hat{\mathbf{x}}$ interface.
In conclusion, the non-zero components of the discontinuity of Chern vectors parallel to the interface correspond to the winding numbers of the surface Fermi loops around the two handles of the surface BZ torus, which completes our vBBC picture for 3D CIs. 

Note, on the contrary, that the symmetry of a Fermi loop in $\mathbf{k}$ is not a universal property and depends on the geometry of the crystal. As discussed in the Supporting Information section 1, 'Symmetry of the Fermi loops', the BBC relations are not modified by an asymmetry of the Fermi loops.

As well, we stress that the 'number' of loops is not the topologically protected quantity. vBBC only establishes a link between the Chern vector discontinuity and the winding of the Fermi loops on the surface BZ, not with their number. For example, two Fermi loop states propagating along $k_y$ and $k_z$, as represented in Fig. 9(b), are topologically equivalent to a single Fermi loop propagating along a tilted direction, as in Fig. 9(c): their winding across the two handles of the BZ is the same, and thus the net propagation direction.  This is important because it means that surface states on the same surface of the slab are allowed to scatter and interfere, changing the counting of closed loops on the BZ.  Due to topological protection, the scattering result is still a closed loop, with the same number of crossing points for a fixed momentum line and same winding across the BZ.

Finally, note that the vBBC conditions derived here can also be applied to diagonal orientations of the interface normal with respect to the Chern vector, provided moving in a new reference frame. In these regards, a detailed explanation of the vBBC for diagonal cuts of the crystal is provided in the Supporting Information section 2, 'Diagonal orientation of the Chern vectors and the interface normal', where a guiding example is provided.

\begin{figure*}[h!]
\centering
\includegraphics[width=178mm]{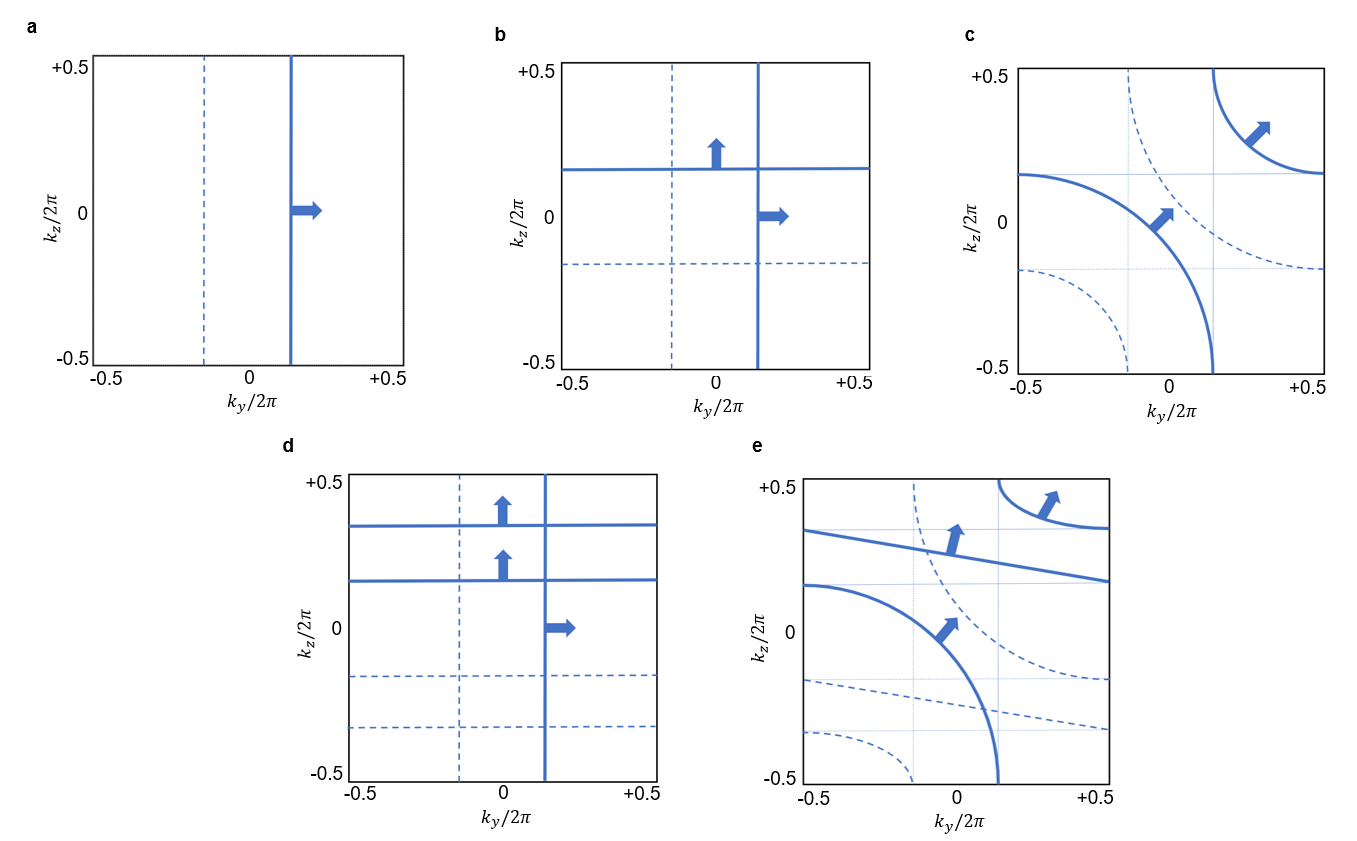}
\captionsetup[figure]{font=small}
\caption{Topology of the Fermi loops and their hybridization. Solid and dashed lines label states residing on opposite side of the sample. Thick and thin lines indicate states before and after hybridization, respectively. a) Fermi loops with $(T_1,T_2)=(0,1)$. b,c) Fermi loops with $(T_1,T_2)=(1,1)$ before and after hybridization. d,e) Fermi loops with $(T_1,T_2)=(2,1)$ before and after hybridization. }
\label{scheme}
\end{figure*}

\begin{center}
\large\textbf{Conclusions}
\end{center}
\normalsize

In this work, we investigated vectorial aspects of bulk-boundary correspondence for photonic 3D CIs with Chern vectors differently oriented in space. 
We showed that, for a 3D CI crystal, the Chern vectors across the interface no longer need to be parallel or anti-parallel to each other, which may render the scalar analogy with 2D difficult to apply. 
We concluded that vectorial features of the BBC need to be taken into account to correctly capture the number and the propagating properties of the photonic surface modes, in perspective of promoting sBBC from 2D to a vBBC for 3D CIs.

Specifically, we analyzed the variety of possible surface states which can emerge on the boundary of photonic 3D CIs, depending on the orientation of Chern vectors across, and with respect to the interface.
We showed that: (1) Any discontinuity in the Chern vector components normal to the plane of the interface does not contribute to the number of surface modes; (2) Multiple unidirectional photonic surface modes arising from different Chern vectors across the interface can hybridizee with each other, preserving their winding around the surface BZ; (3) The number and propagation direction of the surface modes can be related to a difference of Chern vectors parallel to the interface.

Our observations can summarized by via the following statement: the winding  numbers of the surface Fermi loops around the surface BZ correspond to the non-zero components of the discontinuity of Chern vectors parallel to the interface. 
In other words, we derived a link between a bulk topological quantity, the Chern vector, and a boundary observable, the winding of the Fermi loops on the surface, which completes the vBBC picture for 3D CI photonic crystals.

\subsection{Open research lines}
The present work remains open to new developments and research lines: First, as already noted in the previous sections, a 3D CI can be thought of as a stack of 2D CIs from a layer construction point of view \cite{xu2020high,elcoro2020magnetic}. An obstruction in this layer structure across the interface can give rise extra chiral boundary modes \cite{wieder2020axionic, olsen2020gapless}, even without a discontinuity in the Chern vector \cite{wieder2020axionic}, beyond what expected from a simple extension of 2D sBBC to 3D. The concept of obstructed 3D Chern Insulators (o3D CI) and their chiral photonic hinge modes will be explored in forthcoming work\cite{devescovi2022hinge};

Finally, our analysis was performed using 1D linear supercells, due to numerical limitations. However, more complex supercells and Chern vectors configurations may be worth to be considered (2D rod-, 3D core- geometries made of 3D CIs, etc.). An interesting design could consist of 3D CI arranged around a trivial core with Chern vectors pointing inwards. Such a configuration, theoretically proposed in the electronic context by Ref.~\cite{vanderbilt2018berry}, to act as a magneto-electrical (ME) coupler, could be of potential applicative interest as an alternative route to Chern wrappers \cite{mogi2017tailoring,mogi2017magnetic,wang2015quantized}. 
\begin{center}
    \large\textbf{Acknowledgments}
\normalsize
\end{center}
A.G.E. and C.D. acknowledges support from the Spanish Ministerio de Ciencia e Innovación (PID2019-109905GA-C2) and from Eusko Jaurlaritza (IT1164-19, KK-2019/00101 and KK-2021/00082). M.G.D. and M.G.V. acknowledge   the   Spanish   Ministerio  de  Ciencia  e  Innovacion  (grant PID2019-109905GB-C21). A.G.E. and M.G.V. acknowledge funding from Programa Red Guipuzcoana de Ciencia, Tecnología e Innovación 2021 (Grant Nr. 2021-CIEN-000070-01. Gipuzkoa Next) from the Basque Government's IKUR initiative on Quantum technologies (Department of Education). The work of B.B. is supported by the Air Force Office of Scientific Research under award number FA9550-21-1-0131. C.D. acknowledges financial support from the MICIU through the FPI PhD Fellowship CEX2018-000867-S-19-1. The work of J.L.M. has been supported by Spanish Science Ministry grant PGC2018-094626-BC21 (MCIU/AEI/FEDER, EU) and Basque Government grant IT979-16.

 \begin{center}
\large\textbf{Supplementary Information}
\end{center}
\normalsize

\subsection{1. Symmetries of the Fermi loops}
All the Fermi loops displayed in Figs. 1-3 of the main text have reflection symmetry with respect to a line in the surface momentum space. However, we want to stress that the symmetry/asymmetry of the Fermi loops in $\mathbf{k}$ is not a universal property and depends on the geometry of the crystal, as we show here with a clarifying example. Fig. S1 displays the Fermi loops of the $C_z/C_x$ configuration explored in the main text in section 4.2 (panel a), compared with the Fermi loops of a topologically equivalent system, but on which $y$-mirror symmetry is geometrically broken (panel b). Specifically, the crystals in panel b are supercell modulated along $y$ via a $sin$-like modulation, in contrast to the $cos$-like modulation employed in the system sustaining the Fermi loops of panel a.

\begin{figure}[h!]
\centering
\includegraphics[width=100mm]{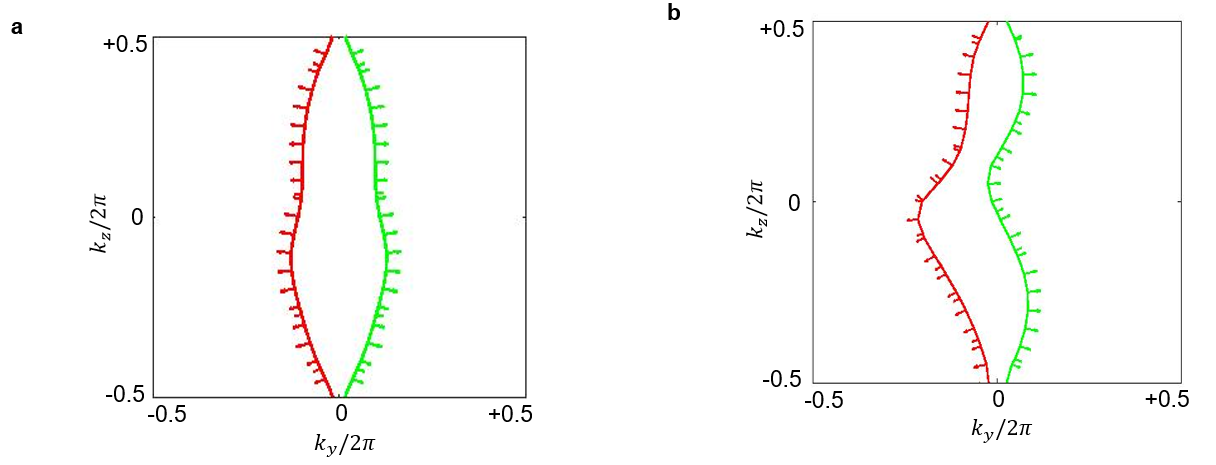}
\captionsetup[figure]{font=small}
\setcounter{figure}{0}
\renewcommand{\thefigure}{S\arabic{figure}}
\caption{Universality of the Fermi loop winding and non-universality of their symmetry.}
\label{s5}
\end{figure}

As it can be observed comparing both panels, in the system with a $sin$-like modulation y the $y$ direction, the $y$-mirror symmetry of the original Fermi loops is broken but the connectivity, the winding, and the topology of the Fermi loops are unaffected. 
Therefore, although the winding of the Fermi loops is not affected by the real space geometrical symmetries, their shape can be altered by them. 
Nevertheless, the BBC relations are not modified by an asymmetry of the Fermi loops.

\subsection{2. Diagonal orientation of the Chern vectors and the interface normal }
In this section, we discuss how to apply vBBC for Chern vectors, in the case of oblique orientation of the surface cut of a 3D cube with respect to the Chern vector, i.e. when the Chern vector is neither parallel nor perpendicular to the boundary. Specifically, we clarify  which diagonal cuts do not break topological protection and how to apply vBBC in that case.
This question is relevant since the topological invariants of a 3D Chern insulator are only defined on its lower-dimensional surfaces and some constraints exist on the planes which can support unidirectional boundary states \cite{halperin1987japan,xu2020high,elcoro2020magnetic,vanderbilt2018berry,xu2015quantum,liu2016effect,jin2018three,kim2018three,tang2019three}. 
In this sense, 3D CIs have some similarities with weak 3D quantum spin Hall systems \cite{fu2007topological,moore2007topological,roy2006three}, which are a 'weak' topological phase and rely on translation along some preferential direction. However, 3D CIs are more robust than this, since no local perturbation on an edge can gap their surface states. Indeed, as we show now, for a 3D CI it is possible to find some diagonal cuts, where the topological protection of surface modes is preserved. In such a case, vBBC can be easily applied provided moving to a new reference frame. We show this with a clarifying example. 

Consider a cut with normal vector given by $(m,n,\ell)$ and Chern vector given by $(0,0,1)$ for simplicity. Since the surface normal is an integer number of lattice vectors, we can go to a new coordinate system $e_1=(m,n,\ell), e_2=(-n,m,0), e_3=(-\ell m, \ell n, m^2+n^2)$, such that the boundary normal vector is $(1,0,0)$. In this new coordinate system (i.e. the new reciprocal lattice vectors), the Chern vector is given by $(\ell, 0 ,m^2+n^2)$. Therefore, we know there will surface states since the Chern number change has a component perpendicular to the boundary, according to Eq.5 of the main text. This means that it is possible to find some oblique planes, an integer linear combination of reciprocal lattice vectors, where the topological protection of unidirectional surface states is not lost. Moreover, it shows that also for an oblique cut, vBBC can be easily applied, by simply moving to the new coordinate frame. 

This important point is related to the fact that a 'strong' 3D Chern insulating phase cannot be topologically defined, as indicated by a zero entry in the A class of the AZ classification table \cite{altland1997nonstandard} and some constraints exist on the surfaces which can support unidirectional boundary states. 
\end{document}